\documentclass{article}

\usepackage{microtype}
\usepackage{graphicx}
\usepackage{subfigure}
\usepackage{booktabs} 
\usepackage{url}

\usepackage{hyperref}

\usepackage[accepted]{icml2018}

\icmltitlerunning{Deep learning for singing processing}

\begin{document}

\twocolumn[
\icmltitle{Deep Learning for Singing Processing: \\ 
Achievements, Challenges and Impact on Singers and Listeners}



\icmlsetsymbol{equal}{*}

\begin{icmlauthorlist}
\icmlauthor{Emilia G\'omez}{jrc,upf}
\icmlauthor{Merlijn Blaauw}{upf}
\icmlauthor{Jordi Bonada}{upf}
\icmlauthor{Pritish Chandna}{upf}
\icmlauthor{Helena Cuesta}{upf}
\end{icmlauthorlist}

\icmlaffiliation{jrc}{Joint Research Centre, European Commission, Seville}
\icmlaffiliation{upf}{Music Technology Group, Universitat Pompeu Fabra, Barcelona}

\icmlcorrespondingauthor{Emilia G\'omez}{emilia.gomez@upf.edu}

\icmlkeywords{Machine Learning, music, deep learning, singing synthesis}
\vskip 0.3in
]



\printAffiliationsAndNotice{\icmlEqualContribution} 

\begin{abstract}
This paper summarizes some recent advances on a set of tasks related to the processing of singing using state-of-the-art deep learning techniques. We discuss their achievements in terms of accuracy and sound quality, and the current challenges, such as availability of data and computing resources. We also discuss the impact that these advances do and will have on listeners and singers when they are integrated in commercial applications. 
\end{abstract}

\section{Introduction}

Singing voice has traditionally been a challenging instrument to analyze and synthesize given its expressive character and the variety of timbre and expressive resources that characterize styles and singers \cite{sundberg1987science}.  However, recent data-driven machine learning techniques, specially deep learning, have substantially boosted the quality and accuracy of singing processing techniques.

In this paper, we illustrate recent advances in deep learning techniques for the analysis, processing and synthesis of singing, comparing them with alternative approaches. Then we study the challenges that these techniques present regarding data and computing infrastructures. We finally discuss on the potential impact that this boost of performance in algorithms may have in the way humans appreciate singing and sing themselves in amateur and professional contexts. 

\section{Overview}
We focus on three well-known and complementary tasks: singing pitch estimation, singing separation from a mix and singing synthesis. Techniques designed for these tasks can also be combined or used as tools in other tasks, e.g. singer replacement (separation plus synthesis) or automatic song translation (synthesis of the same melody and vocal timbre with different phonemes).

\subsection{Singing pitch estimation}
Singing pitch estimation may refer to either monophonic, i.e. one singer, or predominant pitch estimation, i.e. a singer with musical accompaniment.
Deep learning techniques are gaining popularity in this task, and some models such as CREPE \cite{crepe} provide comparable results compared to traditional DSP techniques and heuristics, such as YIN \cite{yin}, pYIN \cite{pyin} or MELODIA \cite{melodia}.\\ 
%
%
To illustrate this, we evaluate the performance of three state-of-the-art algorithms for pitch estimation on the iKala dataset \cite{ikala}. Since this dataset was built for source separation applications, it contains the vocals separated from the accompaniment; therefore, we use the isolated vocals for monophonic and the mix for predominant pitch estimation. Although not specifically designed for this type of signals, we also evaluate CREPE for predominant vocals as a data-driven state-of-the-art monophonic pitch tracking method with pitch accuracies up to 97\% according to the authors' evaluation. The results, following MIREX guidelines for melody extraction evaluation\footnote{\url{http://www.music-ir.org/mirex/wiki}}, displayed in Table \ref{tab:pitch_estimation_results}, show that CREPE and pYIN (considered as a state-of-the-art monophonic pitch estimation method) obtain similar results. We observe the same trend in predominant singing pitch estimation, where CREPE shows similar accuracy than MELODIA. 
%
\begin{table}[t]
\caption{Pitch estimation results with the iKala dataset. Raw pitch accuracy (RPA) is the proportion of melody frames for which the estimation is correct (within $1/2$ semitone).}
\label{tab:pitch_estimation_results}
\vskip 0.15in
\begin{center}
\begin{small}
\begin{sc}
\begin{tabular}{ccc}
\toprule
Algorithm & & Raw Pitch Accuracy  \\ 
\midrule
CREPE & Monophonic & 90.5\% \\ 
pYIN & Monophonic & 91\%  \\ 
CREPE & Predominant &  81.5\% \\ 
MELODIA & Predominant & 81.1\% \\ 
\bottomrule
\end{tabular}
\end{sc}
\end{small}
\end{center}
\vskip -0.1in
\end{table}
\subsection{Singing separation}

The problem of source separation is well studied, and specific metrics have been defined with regard to the quality of separation of one signal from a mixture of signals \cite{vincent2006performance}. 
Singing voice separation, a sub-branch of audio source separation, refers to the idea of isolating the signal related to the singing voice from a musical mixture. It is a distinct task from speech source separation owing to the fact that vowel durations are exaggerated during singing, thus leading to a more harmonic structure to the signal than is the case in speech. This harmonic structure, in conjunction with the melodic pitch range, often clashes with the harmonic structure of the background music, as all melodic instruments try to play within a limited set of harmonic pitched. This is not usually the case with speech signals needing to be separated from background noise, which might not be correlated to the speech signal. Over the last two decades, several techniques using algebraic methods for factorization, such as Non-Negative Matrix Factorization (NMFs) and Independent Component Analysis (ICA) have been used to separate the audio mixture signal into constituent factor signals. One particularly clever and relatively simple technique used for singing voice separation exploits the repetitions commonly seen in musical background signals, which distinguishes them from the non-repeating vocal section \cite{rafii2013repeating}. However, in the last few years, techniques using deep neural networks (DNNs), in the form of Recurrent Neural Networks (RNNs) and Convolutional Neural Networks (CNNs) have raised the bar in terms of state-of-the-art performance for both separation quality \cite{nugraha2016multichannel,uhlich2015deep,grais2016single} and processing time \cite{chandna2017monoaural} and resources required. The Sisec campaign\footnote{\url{https://sisec.inria.fr/}}, which aims at community based signal separation evaluation, has a section devoted to musical source separation. The evaluation is based on source separation metrics like Signal to Distortion Ratio (SDR), Source to Interferences Ratio (SIR), Sources to Artifacts Ratio (SAR) and Image to Spatial distortion Ratio (ISR). Results over the recent years show a steep increase in performance with techniques based on deep learning. MIREX also has a section devoted to singing voice separation, which also shows the same trends in performance, with state-of-the-art methodologies all based on deep learning. 

%

%
%
\subsection{Singing synthesis}
The task of singing synthesis consists of mapping a musical score with lyrics to a sung acoustic signal, generally by modeling a specific singer's voice. Following recent trends in text-to-speech (TTS) research, methods based on deep learning have also been proposed to tackle singing synthesis. These approaches include feedforward networks \citep{NishimuraM2016}, and autoregressive convolutional networks \citep{BlaauwM2017}. In listening tests, these methods have been shown to be able to achieve a sound quality on-par or exceeding the previous state of the art, concatenative methods \citep{BonadaJ2007,BonadaJ2016,ArdaillonL2017Ch3}. A notable limitation of concatenative methods is they require a high degree of coherence between the to be concatenated segments in order to achieve high quality results. This means that rather than natural songs, specially designed recordings and a significant amount of manual intervention are generally needed. As a result, in concatenative methods, fundamental frequency (F0) and timing generally come from some external source \citep{UmbertM2013,ArdaillonL2016}. This in contrast to deep learning approaches, which can jointly model timbre and F0 from a set of natural songs. Other machine learning approaches, such as those based on Hidden Markov Models (HMMs) \citep{SainoK2006,OuraK2010,NakamuraK2014,PucherM2016,LiX2016}, offer similar flexibility in this aspect, but tend to have other drawbacks. Notably, HMM-based methods tend to have a series of limitations inherent to the model itself, such as modeling each phoneme as a small number of discrete states with constant statistics, and many others. The principal result of these limitations is that the resulting sound quality tends to be below the other methods, in particular suffering from oversmoothing. Connectionist deep learning approaches are very general models and have nearly none of such assumptions and inherent limitations to how the model is defined. Furthermore, deep learning approaches tend to allow for much faster experimentation compared to older HMM-based methods. This is in part due to being trained by backpropagation rather than expectation-maximization, but also because of the superior tooling support offered by modern deep learning frameworks (automatic differentiation, GPU training, and so on).

\section{Challenges}

\subsection{Data requirements}

While data-driven deep learning techniques have pushed the envelope in terms of performance in singing processing tasks, this performance is limited by the amount and quality of data available for training and testing. Existing open datasets are varied in terms of singers and style, they often focus on a single singer and they still do not fully cover the wide range of traits of the human singing voice. 

In addition, there is no agreed-upon methodology to document and evaluate datasets in terms of coverage, e.g. with respect to singer, gender, race, language (phonemes) or singing style (e.g. pop, flamenco). Another important issue in the creation and sharing of singing datasets relates to privacy, as singers can now be recognized both in the dataset and the resulted synthesis or processing. In addition, each addressed task has a different set of requirements for the available data. 

In the case of singing voice separation, this necessitates the presence of a stream devoted to the singing voice and one for the accompaniment. The Sisec campaign provides one such dataset, in the form of the MUS 2018 dataset. \cite{musdb18} 
This dataset builds on similar datasets like MSD100 and DSD100, which have $100$ professionally mixed songs each, along with separate stems for vocals, bass, drums and other instruments. The MUS 2018 dataset consists of 150 professionally mixed recordings, with a train set comprising of $100$ songs and a test set of $50$ songs. The iKala dataset, \cite{chan2015vocal} which contains $252$, $30$ second tracks of vocal and backing track music, was, until recently, one of the most popular public datasets used for singing voice separation, as the ground truth vocal track was present without non-linear effects. The iKala dataset also has hand-annotated MIDI-note pitch annotations, which allow it to be used as a baseline for evaluating predominant pitch extraction algorithms. Another such dataset is the ccmixter vocal separation database. \cite{liutkus2015scalable}
While these datasets were curated for the specific purpose of singing voice source separation, there are some datasets created for other tasks such as automatic transcription, which have vocal stem tracks and can be used for singing voice separation. These include the MedleyDB \cite{bittner2014medleydb}, the RWC Music Database \cite{goto2002rwc}, the MIR-1K dataset \cite{hsu2010improvement} and the IRMAS dataset\cite{bosch2012comparison}. 

Singing voice synthesis generally requires a dataset of high quality acapella recordings of a single singer. The dataset should also include the corresponding musical score with lyrics as a minimum, and ideally also include phoneme-level annotations. At present, very few of such datasets are available publicly. The NIT-SONG070-F001\footnote{\url{http://hts.sp.nitech.ac.jp/archives/2.3/HTSdemo_NIT-SONG070-F001.tar.bz2}} is one such dataset, recorded by the Nagoya Institute
of Technology (Nitech). The public version of this dataset contains $31$ (out of the original $70$) studio quality of recordings of Japanese nursery rhymes, and includes phoneme and note level annotations.

Apart from these, there are some datasets constructed for other singing based MIR tasks. These include the MTG-QBH \cite{salamon2013tonal}, Tunebot \cite{cartwright2012building} and SING! \cite{smith2013correlation} datasets, which are useful for tasks like query by humming and singer identification as they include recordings of different singers singing the a finite set of songs, with repetitions. There are also datasets for analysis of intonation trajectories in singing like \cite{dai2015analysis} and \cite{umbert2006spanish}. Other singing related datasets include TONAS for flamenco singing \cite{mora2010characterization}, JAMENDO for singing voice detection \cite{ramona2008vocal} and UltraStar for analyzing emotions and singer traits \cite{schuller2012automatic}.

%

\subsection{Computing resources}

In addition to data, computing resources has also become a factor in the success and accuracy of methods. Deep learning algorithms have a high computational cost for training compared to traditional approaches. Contrary to data, it is difficult to contrast and measure computational cost, as the cost and time required for training is not often mentioned in the literature. 

%
%

\subsection{Explainability, transparency and knowledge}

Traditionally, research on singing processing has also contributed to gain knowledge on the acoustic and expressive resources of singing, styles and singers. However, the black-box character of deep learning techniques makes them not adequate to gain knowledge. The traditional link between acoustic research on singing and engineering development is then weakened, so that developers are not necessary knowledgeable of the acoustic domain under study. We hypothesize that efforts into explainable and transparent algorithms will bring back this link in the future.

\subsection{Towards natural singing synthesis}

In expressive singing we find a huge variety of expressive gestures with different voice qualities. However, so far there is no singing synthesizer able to generate convincing expressive resources such as growls or rough voices. Recently, the introduction in speech synthesis of waveform synthesizers based on deep neural networks (e.g. WaveNet, Tacotron2) have overpassed the sound quality produced by vocoders, and achieved natural sound quality. These techniques can generate convincing speech with different voice qualities, including creaky or fry phonations. Thus, we expect waveform synthesizers applied to singing synthesis might allow generating much more realistic singing.

\subsection{Impact} 

Finally, with the increase in quality of these techniques arise some questions about the impact they may have in singers and listeners when they are integrated in commercial applications, as discussed, not only for music but in different domains, in \cite{gomez2018}. 

For instance, the computational modeling of a given singer or style is currently very good when based on pitch and timbre. This opens up the possibility of correctly identifying, model and imitate anyone's voice. This will affect the usage of automatic recognition systems and the possibility to impersonate any person's singing. 

In addition, the current quality of singing synthesizers might soon make them indistinguishable from human singing. This already has an impact on how musicians and composers interact with those systems. For instance, \textit{Hatsune Miku}, a VOCALOID\footnote{\url{https://www.vocaloid.com}} software voicebank, is an example of new musical and social paradigms facilitated by artificial singing synthesizers. Although it offers new creative opportunities, this kind of systems also raises a debate on the fundamental human experience of singing and listening to other people sing, and how this might change in the future. There is also a more pragmatic debate on how virtual singers will replace humans in some job tasks, in line with existing discussion about the impact of artificial intelligence on the future of work \cite{fernandezmacias2018}. 

Finally, in view of the exploitation of data-driven deep learning techniques in commercial music production scenarios, some intellectual property (IP) matters need to be clarified, as IP seems to be shared between data curators, algorithm developers and singers, who are the original creators of the raw material needed to model human voices.

\section{Conclusions}
We have presented some recent advances on singing processing using deep learning techniques, and we have discussed on the main challenges and potential impact in listeners and singers. We foreseen future work devoted to study this impact and enhance the link between those techniques and the knowledge we have about the human voice. 

\section*{Acknowledgements}
This work has been partially funded by the CASAS (TIN2015-70816-R) Spanish project and TROMPA (H2020 770376) European projects. 

\begin{small}
\bibliography{example_paper}
\bibliographystyle{icml2018}
\end{small}

\end{document}